\documentclass{article}
\usepackage{amsmath,amssymb}
\usepackage{graphicx}

\numberwithin{equation}{section}

\begin{document}

\title{Slowly rotating perfect fluids with a cosmological constant}

\author{
Christian G.~B\"ohmer\footnote{c.boehmer@ucl.ac.uk} ~and
Matthew Wright\footnote{matthew.wright.13@ucl.ac.uk}\\
Department of Mathematics, University College London\\
Gower Street, London, WC1E 6BT, UK
}
\date{\today}

\maketitle

\begin{abstract}
Hartle's slow rotation formalism is developed in the presence of a cosmological constant. We find the generalisation of the Hartle-Thorne vacuum metric, the Hartle-Thorne-(anti)-de Sitter metric, and find that it is always asymptotically (anti)-de Sitter. Next we consider Wahlquist's rotating perfect fluid interior solution in Hartle's formalism and discuss its matching to the Hartle-Thorne-(anti)-de Sitter metric. It is known that the Wahlquist solution cannot be matched to an asymptotically flat region and therefore does not provide a model of an isolated rotating body in this context. However, in the presence of a cosmological term, we find that it can be matched to an asymptotic (anti)-de Sitter space and we are able to interpret the Wahlquist solution as a model of an isolated rotating body, to second order in the angular velocity.
\end{abstract}

\section{Introduction}

It is remarkable that there are still no known exact solutions to Einstein's equations describing a rotating stellar model with physically meaningful equation of state. Such a model is necessary for accurately describing astrophysical objects such as stars, planets, galactic nuclei and galaxies, all of which are rotating in general. The closest to such an exact solution that has been found, is the rotating infinitely thin shell of dust matching to the extremal Kerr solution, given by Neugebauer and Meinel~\cite{Neugebauer:1995pm}, and its generalisation to counter-rotating discs~\cite{Klein:2001hn,Cuchi:2012nm}. However, these models have no physically meaningful interior as the matter source has zero thickness. In particular, despite much effort, there is no global solution describing an interior rotating perfect fluid which matches smoothly to an exterior rotating vacuum solution.  

The slow rotation formalism, first developed by Hartle in 1967~\cite{Hartle:1967he,Hartle:1968si}, has been important in deriving results about the possibility of matching an interior solution to an exterior vacuum. If such a matching is not possible in the slow rotation limit, then the matching will not be possible in general. For example, it was shown in~\cite{Bradley:1999gs} that the Wahlquist metric cannot be matched to an asymptotically flat vacuum exterior. The general slow rotating vacuum exterior metric is known as the Hartle-Thorne metric, in general this metric is not asymptotically flat. It was later shown~\cite{Bradley:2000sj} that, up to second order in the angular velocity of the fluid, the Wahlquist metric can be matched to the Hartle-Thorne exterior, if one drops the requirement of asymptotic flatness. In fact, with this requirement dropped, the matching of any slowly rotating perfect fluid to the Hartle-Thorne exterior is always possible up to second order in the fluids angular velocity~\cite{Bradley:2006eg}.

It is now widely accepted that the universe is accelerating in expansion, which can be explained by the presence of a small positive cosmological constant~\cite{Riess:1998cb,Perlmutter:1998np}. This indicates that the universe is not asymptotically flat, rather asymptotically it behaves like de Sitter space. Therefore, the possibility of matching an interior fluid solution to an asymptotically (anti-)de Sitter vacuum should be investigated and is somewhat more natural. For such an investigation to take place, suitable interior and exterior metrics satisfying Einstein's equations with cosmological constant must be found. Interior solutions can easily be constructed, any perfect fluid solution to Einstein's equations with $\Lambda=0$ and a given equation of state $f(\rho,p)=0$ is easily generalised to a solution to Einstein's equations with $\Lambda \neq 0$ by making the substitutions $p \rightarrow p - \Lambda/(8\pi G)$, and $\rho \rightarrow \rho + \Lambda/(8\pi G)$. This new solution now has the equation of state $f( \rho + \Lambda/(8\pi G), p - \Lambda/(8\pi G))=0$.

This leaves us with the problem of finding a suitable vacuum exterior solution which is one of the motivations of this work. In this paper we work with the slow rotation formalism in the presence of a cosmological constant, and attempt to find the generalisation of the Hartle-Thorne exterior vacuum metric with $\Lambda$, which we call the Hartle-Thorne-(anti)-de Sitter metric. The Hartle-Thorne metric is asymptotically flat only when one of the constants in the metric is set to zero. However, the asymptotic behaviour of the Hartle-Thorne-(anti)-de Sitter metric is considerably different, asymptotically it behaves like (anti)-de Sitter space for any values of the constants in the metric.

Investigating vacuum axisymmetric solutions with a cosmological constant is also important from a mathematical point of view. When $\Lambda=0$ there are an abundance of vacuum axisymmetric solutions, for example the Weyl and Papapetrou classes of solutions~\cite{Islam:1985}. In the case of spherical symmetry, generalising a vacuum solution to a $\Lambda$-vacuum solution is simple, the Schwarzschild metric is easily generalised to the Schwarzschild-de-Sitter metric. However, examples of stationary axially symmetric solutions to Einstein's equations with a cosmological constant are very rare~\cite{Charmousis:2006fx}. Equations that were integrable with $\Lambda=0$ are no longer integrable for general $\Lambda \neq 0$. Our Hartle-Thorne-(anti)-de Sitter metric is an example of a new approximate axisymmetric $\Lambda$-vacuum solution. 

This paper is organised as follows. In Section~\ref{sec:fe} we begin by stating our conventions and deriving the slowly rotating $\Lambda$-vacuum field equations up to second order. In Section~\ref{sec:slow} we find the coordinate transformations required to transform the slowly rotating Kerr-de-Sitter metric into Hartle's coordinates and derive the Hartle-Thorne-(anti)-de Sitter metric. Section~\ref{sec:prop} discusses the properties of this solution and its asymptotic behaviour. Section~\ref{sec:wahl} reviews the Wahlquist metric in the slow rotation formalism, and in Section~\ref{sec:match} we show that the Wahlquist metric can be matched to an asymptotically (anti)-de Sitter metric up to second order in the rotation parameter. We conclude in the last section~\ref{sec:concl}.

\section{Field Equations}
\label{sec:fe}

\subsection{Conventions and field equations}
The Einstein field equations with cosmological term are given by
\begin{align}
  G_{ab} + \Lambda g_{ab} = 8\pi G T_{ab}
\end{align}
where we set $c=1$. The energy-momentum tensor of a perfect fluid is given by
\begin{align}
  T_{ab} = (\mu + p)u_a u_b + p g_{ab}
\end{align}
where $u_a$ is the fluid's 4-velocity, $\mu$ is its energy density and $p$ is its pressure.

Following Hartle~\cite{Hartle:1967he}, we write the stationary, axially symmetric metric in the following form
\begin{align}
  ds^2 &= -H^2 dt^2 + Q^2 dr^2 
  + r^2 K^2 \bigl[d\theta^2 + \sin^2\theta(d\varphi - L dt)^2 \bigr].
\end{align}
Here $H$, $Q$, $K$ and $L$ are functions of $(r,\theta)$. Hartle then makes a slow rotation approximation by expanding this metric to order $\Omega^2$, where $\Omega$ is the angular velocity of the rotating fluid interior. In this paper we follow the notation of~\cite{Bradley:1999gs} and make the following expansion
\begin{multline}
  ds^2 = -A(r)^2(1+2h(r,\theta))dt^2+\frac{1+2m(r,\theta)}{B(r)^2}dr^2\\
  +r^2(1+2k(r,\theta))[d\theta^2+\sin^2\theta(d\phi-\omega(r)dt)^2]+\mathcal{O}(\Omega^3),
  \label{Hartle}
\end{multline}
where $\omega$ is a function of order $\mathcal{O}(\Omega)$ and $h,m$ and $k$ are of order $\mathcal{O}(\Omega^2)$, respectively. These functions are expanded in spherical harmonics
\begin{align}
  h(r,\theta)&=h_0+h_2(r)P_2(\cos\theta),\\
  m(r,\theta)&=m_0+m_2(r)P_2(\cos\theta),\\
  k(r,\theta)&=k_2(r)P_2(\cos\theta),
  \label{kdef}
\end{align}
where $P_2$ is the second Legendre polynomial, and the freedom in the choice of radial coordinate was used to set $k_0(r)=0$. 

In the following we will derive the vacuum field equations at successive orders in the rotation parameter $\Omega$.

\subsection{Zeroth order equations}

To zeroth order in $\Omega$ we have the standard Schwarzschild type metric
\begin{align}
  ds^2=-A(r)^2dt^2+\frac{1}{B(r)^2}dr^2+r^2dS^2,
\end{align}
with $dS^2 = d\theta^2 + \sin^2\theta d\phi^2$. Two independent vacuum field equations are given by
\begin{align}
  \frac{1}{r^2}[1-\frac{d(rB^2)}{dr}] - \Lambda &= 0,\\
  \frac{1}{r^2}[1-\frac{B^2}{A^2}\frac{d(rA^2)}{dr}] + \Lambda &= 0.
\end{align}
Solving these equations gives the well known Schwarzschild-(anti)-de Sitter spacetime
\begin{align}
A(r)^2=B(r)^2=1-\frac{2M}{r}-\frac{\Lambda r^2}{3}.
\end{align}
For positive $\Lambda$ this spacetime has two horizons, the black hole horizon and the cosmological horizon. Their locations are given by the roots of the cubic equation $A(r)=0$, note that the third root is unphysical as $r<0$. For $\Lambda \ll 1$, those horizon are located at $r_{\rm bh} \approx 2m$ and $r_{\Lambda} \approx \sqrt{3/\Lambda}$. For negative $\Lambda$, the metric has only one horizon, the black hole horizon.

\subsection{First order equations}

Let us next consider the field equations up to order $\Omega$. In this case the metric takes the form
\begin{align}
  ds^2=-A(r)^2dt^2+\frac{1}{B(r)^2}dr^2+r^2dS^2 
  -2\omega(r)r^2\sin^2\theta dt d\phi
  \label{first order}
\end{align}
There only is one extra field equation, namely $G^\phi_t=0$ (we note $\Lambda \delta^\phi_t=0$), and it reads
\begin{align}
  \frac{d}{dr}(r^4\frac{B}{A}\frac{d\omega}{dr})+
  4r^3\omega\frac{d}{dr}(\frac{B}{A})=0.
\end{align}
In the vacuum exterior, we know $A=B$ and, irrespective of their functional form, this equation is easily solved to give
\begin{align}
  \omega(r)=\frac{2aM}{r^3}+c_1,
  \label{omega}
\end{align}
where $a$ and $C_1$ are constants of integration. The constant $a$ is first order in $\Omega$ and $aM$ can be identified with the total angular momentum of the spacetime. Without loss of generality we are free to set $c_1=0$ by performing a rigid rotation of our coordinate system.

\subsection{Second order equations}

Using the full second order metric, we can derive 5 new equations for the functions $m_0,m_2,h_0,h_2,k_2$. The $l=0$ and the $l=2$ equations decouple, so can be treated separately.

\subsubsection*{$l=0$ equations}

From the $\theta$ independent part of $G^t_t+\Lambda=0$ and $G^r_r+\Lambda=0$  we can derive the following equations for $m_0$ and $h_0$
\begin{align}
  12\frac{d}{dr}(rA^2m_0)-r^4(\frac{d\omega}{dr})^2 &= 0,
  \label{m0 equation}\\
  -12m_0\frac{d}{dr}(rA^2)+12r A^2 \frac{dh_0}{dr}+r^4(\frac{d\omega}{dr})^2 &= 0.
  \label{h0eqn}
\end{align}
Equation~(\ref{m0 equation}) can immediately be solved to give
\begin{align}
  m_0=\frac{1}{{(2M-r+\frac{\Lambda}{3}r^3)}}(\frac{a^2M^2}{r^3}-c_2)
\end{align}
and substituting this into~(\ref{h0eqn}) allows us to solve for $h_0$
\begin{align}
  h_0=-\frac{1}{{(2M-r+\frac{\Lambda}{3}r^3)}}(\frac{a^2M^2}{r^3}-c_2).
\end{align}
Here $c_2$  is a constant of integration of order $\mathcal{O}(\Omega^2)$. 

\subsubsection*{$l=2$ equations}

Due to the nature of the problem, the rest of this section is very equation heavy. We aim to give enough detail so that our calculations can be followed and verified. As expected, our equations reduce to those of~\cite{Hartle:1967he} when $\Lambda$ is set to zero.

From the $G^r_\theta=0$ equation we have
\begin{align}
  r\frac{d}{dr}(h_2+k_2)+\frac{r}{A}\frac{dA}{dr}(h_2-m_2)-(h_2+m_2)=0. 
  \label{2o1}
\end{align}
Using the $\theta$ dependent part of $G^r_r=G^\theta_\theta$ gives
\begin{multline}
  2r\frac{B^2}{A}\frac{dA}{dr}(r\frac{dk_2}{dr}-m_2)
  -2r^2Bh_2\frac{d}{dr}(\frac{B}{r}) \\
  +m_2-4k_2-5h_2-\frac{1}{3}r^4\frac{B^2}{A^2}(\frac{d\omega}{dr})^2=0 
  \label{204}.
\end{multline}
Finally, the $G^\theta_\theta=G^\phi_\phi$ equation gives
\begin{align}
  6(h_2+m_2)-r^4(\frac{B}{A})^2(\frac{d\omega}{dr})^2+4r^3\omega^2\frac{B}{A}\frac{d}{dr}(\frac{B}{A})=0.
\end{align}
This implies that in the vacuum exterior we have the result
\begin{align}
  h_2+m_2=\frac{6a^2M^2}{r^4}.
\end{align}

Substituting this result into our zeroth and first order results, and into~(\ref{2o1}) gives the equation
\begin{align}
  r(h_2'+k_2')+\frac{2M-2\frac{\Lambda}{3} r^3}{r-2M-\frac{\Lambda}{3} r^3}h_2
  -\frac{r-M-2\frac{\Lambda}{3} r^3}{r-2M-\frac{\Lambda}{3} r^3}(\frac{6a^2M^2}{r^4})=0,
  \label{eq1}
\end{align}
while~(\ref{204}) gives us the equation
\begin{multline}
  (2M-2\frac{\Lambda}{3} r^3)k_2'-4(1+\frac{M}{r}+\frac{1}{2}\frac{\Lambda}{3} r^2)h_2\\
  -4k_2-(1+\frac{2M}{r}-2\frac{\Lambda}{3} r^2)\frac{6a^2M^2}{r^4}=0.
  \label{eq2}
\end{multline}

When $\Lambda=0$ we recover the solutions of~\cite{Bradley:1999gs} which are
\begin{multline}
  h_2=3c_1r(2M-r)\log(1-\frac{2M}{r})+a^2\frac{M}{r^4}(M+r)\\
  +2c_1\frac{M}{r}(3r^2-6Mr-2M^2)\frac{r-M}{2M-r}+(1-\frac{2M}{r})r^2q_1,
  \label{Br1}
\end{multline}
\begin{multline}
  k_2=3c_1(r^2-2M^2)\log(1-\frac{2M}{r})-a^2\frac{M}{r^4}(2M+r)\\
  -2c_1\frac{M}{r}(2M^2-3Mr-3r^2)+(2M^2-r^2)q_1.
  \label{Br2}
\end{multline}

\section{Slowly rotating solutions}
\label{sec:slow}

\subsection{Slowly rotating Kerr-de-Sitter space}

There are very few known exact rotating vacuum solutions in the presence of a cosmological constant. A well known solution is the Kerr-de-Sitter metric, first discovered by Carter in~\cite{Carter:1968}. 

Here we start by transforming the Kerr-de-Sitter metric into Hartle's original coordinates. This enables us to find a particular solution to the equations given in the previous section. The Kerr-de-Sitter metric in Boyer-Lindquist coordinates is given by
\begin{multline}
  ds^2=-\frac{\Delta_l}{\Xi_l^2 \rho_l^2}(dt-a \sin^2\theta d\phi)^2
  +\frac{\Theta_l \sin^2\theta}{\Xi_l^2 \rho_l^2}(adt-(r^2+a^2)d\phi)^2\\
  +\frac{\rho_l^2}{\Delta_l}dr^2+\frac{\rho_l^2}{\Theta_l}d\theta^2
  \label{kerrbl}
\end{multline}
where
\begin{alignat}{2}
  \rho_l &= \sqrt{r^2+a^2 \cos^2\theta} &\qquad l&=\sqrt{\frac{3}{\Lambda}} \\
  \Theta_l&=1+\frac{a^2}{l^2}\cos^2\theta &\qquad \Xi_l &= 1+\frac{a^2}{l^2}
\end{alignat}
\begin{align}
  \Delta_l=-\frac{r^2}{l^2}(r^2+a^2)+r^2-2Mr+a^2
\end{align}
Expanding the metric~(\ref{kerrbl}) to first order in $a$ we arrive at
\begin{multline}
  ds^2 = (1-\frac{2M}{r}-\frac{\Lambda r^2}{3})dt^2
  +\frac{1}{1-\frac{2M}{r}-\frac{\Lambda r^2}{3}}dr^2 \\
  +r^2(d\theta^2+\sin^2\theta d\phi^2)
  -2a(\frac{6M+r^3\Lambda}{3r})\sin^2\theta dt d\phi.
\end{multline}

By looking at the $g_{t\phi}$ component of this metric, we see that, unlike the Kerr metric, the Kerr-de-Sitter metric in Boyer-Lindquist coordinates is asymptotically rotating. Hence we first need to apply a coordinate transformation to adjust to an asymptotically non-rotating frame. This can be done by performing a rigid rotation of the coordinate system                                                                                                                                                   
\begin{align}
  \phi \rightarrow \tilde{\phi} + a \frac{\Lambda}{3} t,
\end{align}
which gives
\begin{multline}
  ds^2 = (1-\frac{2M}{r}-\frac{\Lambda r^2}{3})dt^2
  +\frac{1}{1-\frac{2M}{r}-\frac{\Lambda r^2}{3}}dr^2 \\
  +r^2(d\theta^2+\sin^2\theta d\tilde{\phi}^2)
  -2a(\frac{6M}{3r})\sin^2\theta dt d\tilde{\phi}.
\end{multline}

Next, we expand the Kerr-de-Sitter metric to second order in $a$. To put this metric into Hartle's form we make the following coordinate transformations
\begin{align}
  r &\rightarrow \tilde{r} \Bigl(1-\frac{a^2}{2\tilde{r}^2}\Bigl[(1+\frac{2M}{\tilde{r}}-\frac{\Lambda}{6} \tilde{r}^2) (1-\frac{M}{\tilde{r}})
  \nonumber \\
  &\quad-\cos^2\tilde{\theta}(1-\frac{2M}{\tilde{r}}-\frac{\Lambda}{3} \tilde{r}^2)
  (1+\frac{3M}{\tilde{r}})\Bigr]\Bigr),\\
  \theta &\rightarrow \tilde{\theta}-a^2\cos\tilde{\theta}\sin\tilde{\theta}\frac{1}{2\tilde{r}^2}(1+\frac{2M}{\tilde{r}}
  -\frac{\Lambda \tilde{r}^2}{3}),
\end{align}
where we have corrected for the sign error given in Hartle and Thorne's original paper~\cite{Hartle:1968si}. Finally we have to rescale the radial coordinate $\tilde{r}$ in order to set $k_0(\tilde{r})=0$ in the new coordinates, see our remark after Eq.~(\ref{kdef}). This rescaling is given by 
\begin{align}
  \tilde{r} \rightarrow 
  \frac{3a^2\Lambda M+\sqrt{3(48\hat{r}^2-8a^2\hat{r}^2\Lambda+3a^2M^2\Lambda^2)}}{2(6-a^2\Lambda)}.
\end{align}
We note immediately that in case of vanishing cosmological term $\Lambda = 0$, this rescaling becomes trivial and reduces to $\tilde{r} \rightarrow \hat{r}$. Applying these transformations, our new metric is now given in the coordinates $(t,\hat{r},\tilde{\theta},\tilde{\phi})$. For convenience we now drop the hat and tildes. To the best of our knowledge, the above coordinate transformation has not been given previously in the literature.

In these new coordinates, we find the following expressions for Hartle's functions
\begin{align}
  k_2(r) &= \frac{-a^2M(2M+r)}{r^4}-\frac{a^2M\Lambda}{3r},
  \label{Kerr1} \\
  m_0(r) &= \frac{a^2M(M+2\frac{\Lambda}{3}r^3)}{r^3(2M-r+\frac{\Lambda}{3}r^3)},
  \label{Kerr2} \\
  m_2(r) &= \frac{a^2M(5M-r)}{r^4}+\frac{a^2M\Lambda}{3r},
  \label{Kerr3} \\
  h_0(r) &= \frac{-a^2(M^2+4M\frac{\Lambda}{3}r^3+\frac{\Lambda}{3}r^4(\frac{\Lambda}{3}r^2-1))}
  {r^3(2M-r+\frac{\Lambda}{3}r^3)},
  \label{Kerr4} \\
  h_2(r) &= \frac{a^2M(M+r)}{r^4}-\frac{a^2M\Lambda}{3r}. 
  \label{Kerr5}
\end{align}
These agree with the solutions for the slowly rotating Kerr spacetime when $\Lambda \rightarrow 0$. It is also easily checked that they satisfy Hartle's equations derived in the previous section up to second order.

\subsection{$l=2$ solution}

We have found a particular solution to Hartle's equations with cosmological constant. However, we are interested in finding further solutions. We have already found the general solution to the $l=0$ equations, so now we study the $l=2$ equations. We start with the following ansatz
\begin{align}
  h_2 &= \frac{a^2M(M+r)}{r^4}-\frac{a^2M\Lambda}{3r} + H_2(r)
  \label{inhoma}\\
  k_2 &= \frac{-a^2M(2M+r)}{r^4}-\frac{a^2M\Lambda}{3r} + K_2(r).
  \label{inhomb}
\end{align}
Substitution of this into~(\ref{eq1}) and~(\ref{eq2}) yields a system of first order differential equations in $H_2$ and $K_2$ which is given by
\begin{align}
  K_2' +H_2' &= \frac{2 H_2 (3M - \Lambda r^3)}{r (6M - 3r +\Lambda r^3)}\\
  K_2' &= \frac{6K_2}{3M - \Lambda r^3} + 
  \frac{2 H_2 (6M + 6r + \Lambda r^3)}{r (3M - \Lambda r^3)}.
  \label{inhom}
\end{align}
It turns out that in general we cannot find an analytical solution to this system of equations. As stated above, when $\Lambda = 0$ one can find analytical solutions to~(\ref{inhom}). Curiously, when $M=0$ and $\Lambda \neq 0$ one can also find solutions explicitly, however, these are more difficult to interpret in the context of the Hartle formalism. 

We should note that we can always study solutions numerically. For given initial values $H_2(r_i)$ and $K_2(r_i)$ at some radius $r_i$ the equations can be integrated provided the right-hand sides remain regular. The singularity we may encounter when integrating numerically will correspond to the black hole event horizon and the cosmological horizon where the coordinates break down. 

\section{Properties of solutions}
\label{sec:prop}

In this section we will take a close look at the solutions of~(\ref{inhom}) and in particular study their behaviour for large radii. Firstly, equations~(\ref{inhom}) are solved numerically for different parameter choices. Secondly, we solve the differential equations by expanding in powers of $1/r$ and show that the numerical and analytical results match.  

\subsection{Numerical solution}
\label{sec:num}

For given values of $M$ and $\Lambda$ and given initial conditions $H_2(r_i)$ and $K_2(r_i)$ at some radius $r_i$, it is well know that a solution exists to the system of equations~(\ref{inhom}). This solution exists for all $r > r_i$ provided that the right-hand sides regular. We can reduce the number of constant by introducing the new independent variable $x = r/M$ and a new constant $\lambda = M^2 \Lambda$ in~(\ref{inhom}). This gives
 \begin{align}
  \frac{dK_2}{dx} + \frac{dH_2}{dx} &= \frac{2 H_2 (3 - \lambda x^3)}{x (6 - 3x + \lambda x^3)}\\
  \frac{dK_2}{dx} &= \frac{6K_2}{3 - \lambda x^3} + 
  \frac{2 H_2 (6 + 6x + \lambda x^3)}{x (3 - \lambda x^3)}.
  \label{inhom_n}
\end{align}
This $\lambda$ is in fact quite natural as it is related to the discriminant of the metric of Schwarzschild-(anti)-de Sitter space. The other advantage of working with $x$ and $\lambda$ is that both quantities are also dimensionless. 

\begin{figure}
  \centering
  \includegraphics[width=0.48\textwidth]{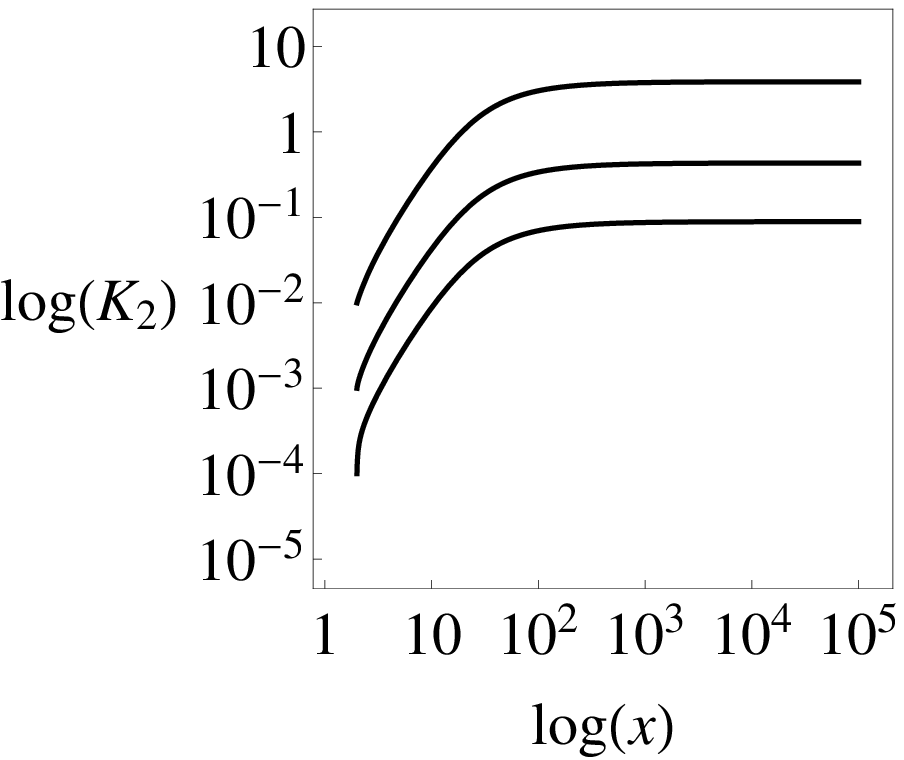}
  \hfill
  \includegraphics[width=0.48\textwidth]{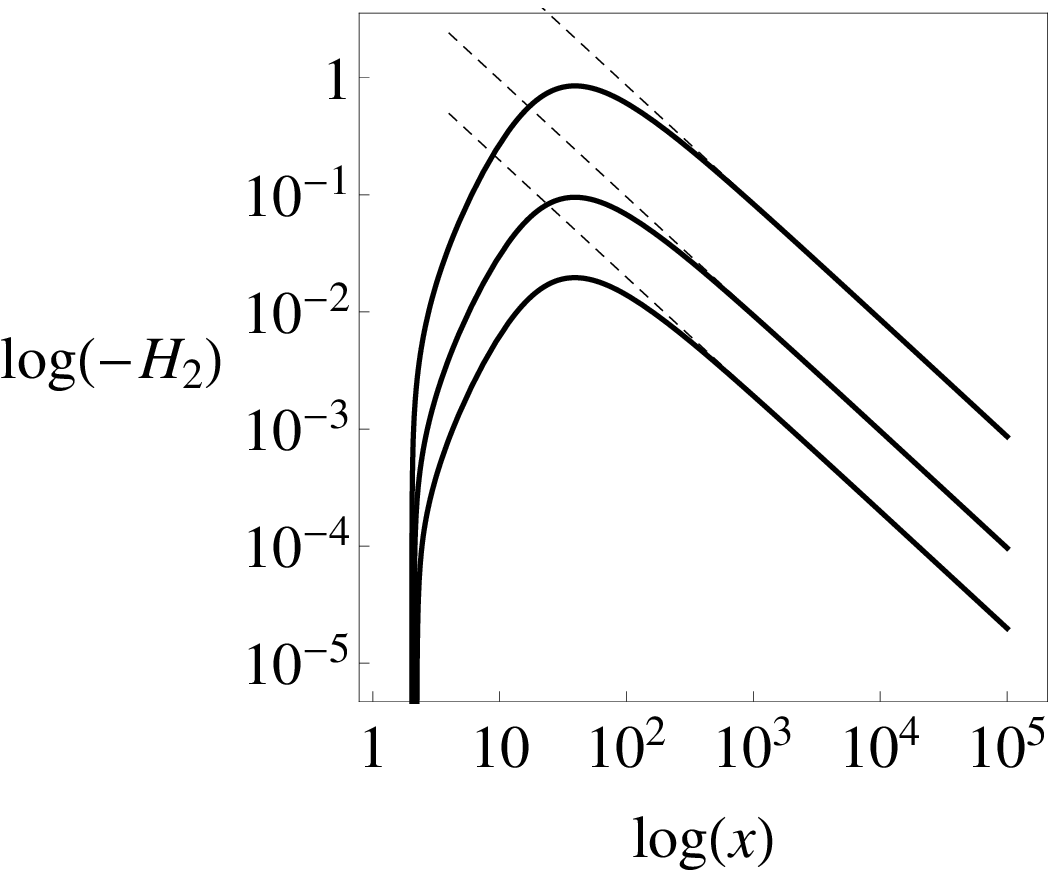}
  \caption{Parameter values are $\lambda = -0.008$ which gives $x_{\rm bh } \approx 1.9793$; initial conditions are $x_i = 2$, $K_2(x_i)=\{10^{-2},10^{-3},10^{-4}\}$, $H_2(x_i)=10^{-3}$. The dashed lines on the right panel show the asymptotes of $\log(-H_2)$.}
  \label{fig:num1}
\end{figure}

In order to avoid numerical issues at the cosmological horizon, we choose $\lambda < 0$ which corresponds to $\Lambda < 0$ and choose $x_i$ to be larger than the corresponding black hole event horizon $x_{\rm rb} = r_{\rm bh}/M$. For this choice of parameters and initial `radius', the right-hand sides of~(\ref{inhom_n}) are regular for all $x > x_i$. Figure~\ref{fig:num1} shows the numerical solutions for some given initial conditions. We display log-log plots since these are convenient when comparing with the analytical results.    
The left panel of Figure~\ref{fig:num1} shows that $\log(K_2)$ approaches a constant values as $x$ becomes large. On the other hand, the right panel shows that $\log(-H_2)$ decreases linearly for large $\log(x)$. The dashed lines in the right panel indicate the straight lines derived from the numerical solution for large $x$. We are in particular interested in the slope of these lines since functions of the form $1/x^n$ become straight lines with slope $-n$ in log-log plots. For the three chosen initial conditions we find
\begin{align}
  n = \{0.999651,0.999639,0.999685\}
\end{align}  
and thus we can confidently conclude that the leading order term in the asymptotic expansion of $H_2$ should be $1/x$ while $K_2$ should approach a constant. 

Varying the value of $\lambda < 0$ or the initial conditions does not change these results. For sake of simplicity we will not discuss the case of a positive cosmological term due to the numerical issues arising near the cosmological horizon. In principle this can be done by introducing coordinates regular at the horizon, however, this is outside the main focus of our work.   

\subsection{The $l=2$ asymptotic solution}

As already stated, we cannot find the analytic solution to the $l=2$ equations. We will now examine the asymptotic behaviour of the functions $H_2$ and $K_2$ for large $r$ and $x$. This means we perform an expansion in powers of $1/r$ up to $\mathcal{O}(1/r^8)$ in the the system of equations~(\ref{inhom}) or~(\ref{inhom_n}). We also expand $H_2$ and $K_2$ as power series in $1/r$ and $1/x$ and determine the coefficients of this expansion by solving~(\ref{inhom}) and~(\ref{inhom_n}) for each power separately.

For $H_2$ we find
\begin{align}
  H_2 &= 
  q_1 \left(\frac{1}{r^2}-\frac{6 M}{\Lambda  r^5}+\frac{9}{5 \Lambda ^2 r^6}
  -\frac{18 M}{\Lambda ^2 r^7}+\frac{216}{35 \Lambda ^3 r^8}
  +\frac{36 M^2}{\Lambda ^2 r^8}\right) 
  \nonumber \\
  &+c_1 \left(\frac{1}{r}-\frac{3}{\Lambda r^3}-\frac{9 M}{\Lambda  r^4}
  -\frac{9 M}{\Lambda ^2 r^6}+\frac{5 M^2}{\Lambda ^2 r^7}
  -\frac{216 M}{7 \Lambda ^3 r^8}\right),
  \label{h2asymptote}
\end{align}
or likewise using $x$ and $\lambda$
\begin{align}
  H_2 &= 
  \tilde{q}_1 \left(\frac{1}{x^2}-\frac{6}{\lambda  x^5}+\frac{9}{5 \lambda^2 x^6}
  -\frac{18}{\lambda^2 x^7}+\frac{216}{35 \lambda^3 x^8}
  +\frac{36}{\lambda^2 x^8}\right) 
  \nonumber \\
  &+\tilde{c}_1 \left(\frac{1}{x}-\frac{3}{\lambda x^3}-\frac{9}{\lambda  x^4}
  -\frac{9}{\lambda^2 x^6}+\frac{5}{\lambda^2 x^7}
  -\frac{216}{7 \lambda^3 x^8}\right).
  \label{h2asymptote_n}
\end{align}
We see that the leading order term in this expansion is $1/r$ or $1/x$, in agreement with the numerical results. Let us next consider $K_2$ for which we arrive at
\begin{align}
  K_2 &= 
  q_1 \left(-\frac{\Lambda }{6}+\frac{3}{2 \Lambda  r^4}+\frac{9}{5 \Lambda ^2 r^6}
  -\frac{36 M}{7 \Lambda^2 r^7}+\frac{243}{70 \Lambda ^3 r^8}\right)
  \nonumber\\
  &+c_1 \left(\frac{1}{r}+\frac{3}{\Lambda  r^3}-\frac{9 M}{\Lambda ^2 r^6}
  -\frac{243 M}{14 \Lambda ^3 r^8}\right)
  \label{k2asymptote}
\end{align}
or equivalently in $x$ and $\lambda$
\begin{align}
  K_2 &= 
  \tilde{q}_1 \left(-\frac{\lambda }{6}+\frac{3}{2 \lambda  x^4}+\frac{9}{5 \lambda^2 x^6}
  -\frac{36}{7 \lambda^2 x^7}+\frac{243}{70 \lambda ^3 x^8}\right)
  \nonumber\\
  &+\tilde{c}_1 \left(\frac{1}{x}+\frac{3}{\lambda  x^3}-\frac{9}{\lambda^2 x^6}
  -\frac{243}{14 \lambda^3 x^8}\right).
  \label{k2asymptote_n}
\end{align}
Here the leading order contribution is a constant, again in agreement with the numerical results.

If we now compare~(\ref{h2asymptote}) with~(\ref{inhoma}) and~(\ref{Kerr1}), and~(\ref{k2asymptote}) with~(\ref{inhomb}) and~(\ref{Kerr5}), it follows that the Kerr-(anti)-de Sitter metric is the particular solution of this expansion with $q_1=0$ and $c_1=0$. We further remark that this is the only solution with a terminating asymptotic expansion.

It should be notes that the expansions~(\ref{h2asymptote}) and~(\ref{k2asymptote}) contain various $1/\Lambda$ terms. Therefore, we cannot consider the limit $\Lambda \rightarrow 0$ directly. This is not surprising since the since the presence of $\Lambda$ changes the characteristics of the differential equations considerably.

In order to study the $\Lambda=0$ case, one must set $\Lambda = 0$ in~(\ref{inhom}) and work directly with the resulting equations. These equations (with $\Lambda = 0$) can be solved analytically, see~(\ref{Br1}) and~(\ref{Br2}). As expected, the asymptotic behaviour of the $\Lambda = 0$ solutions is quite different to the behaviour of the solutions with cosmological term. With $\Lambda$, in general we have $H_2 \rightarrow 0$ and $K_2\rightarrow -q_1\Lambda/6$ as $r\rightarrow\infty$. On the other hand, when $\Lambda = 0$, both functions $H_2$ and $K_2$ are proportional to $r^2$ as $r\rightarrow\infty$. Coincidentally, the terms in the metric with $\Lambda$ are precisely of that same form $\Lambda/3 r^2$. Only if we choose $q_1=0$ does this solution become asymptotically flat. As expected from the presence of the cosmological constant we expect to find (anti)-de Sitter space asymptotically. The following subsection discusses this in more detail.

\subsection{Asymptotic behaviour}

To investigate in more detail the behaviour of this metric at infinity we consider the canonical locally non-rotating Lorentz tetrad considered in~\cite{Bradley:2000sj} which is given by
\begin{align}
  e_0 &= (1-h+\frac{r^2}{2A^2}\omega^2 \sin^2\theta)
  \frac{1}{A}\frac{\partial }{\partial t}\\
  e_1 &= (1-m)A\frac{\partial}{\partial r} \\
  e_2 &= (1-k)r^{-1}\frac{\partial}{\partial \theta} \\
  e_3 &= -\frac{r}{A^2}\omega \sin\theta \frac{\partial}{\partial t}-
  (1-k+\frac{r^2}{2A^2}\omega^2\sin^2\theta)\frac{1}{r \sin\theta}
  \frac{\partial}{\partial \phi}.
\end{align}
In terms of this tetrad~\cite{Fodor:1998jp}, the gravitoelectric fields $E_i$ and the gravitomagnetic fields $H_i$ are defined in terms of the following components of the Riemann tensor
\begin{align}
  E_1 & = R_{0101}-\frac{\Lambda}{3} & \quad 
  E_2 & = R_{0202}-\frac{\Lambda}{3} & \quad 
  E_3 & =R_{0102}\\
  H_1 & = R_{0123} & \quad 
  H_2 & = -R_{0213} & 
  \quad H_3 &= R_{0223}.
\end{align}
We see, using our asymptotic expansion for $k_2$ and $h_2$, that both the gravitoelectric and gravitomagnetc fields of the Hartle-Thorne-(anti)-de Sitter metric tend to zero as $r \rightarrow \infty$. This happens for all values of the constants $c_1$ and $q_1$, and hence we can conclude that the Weyl tensor of the spacetime vanishes there.

Definitions of a spacetime being asymptotically (anti)-de Sitter are given in~\cite{Ashtekar:1984zz,Ashtekar:2014zfa}, and we follow the notation given there. We can choose the conformal factor
\begin{align}                                                               
  \tilde{\Omega} = \frac{f(\theta)}{r} = 
  \frac{1}{r}\left(\sqrt{1-\frac{\Lambda}{3}q_1 P_2(\cos\theta)}\right)^{-1},
\end{align}
and also choose $\Lambda$ such that $|q_1 \Lambda|<3$. Since $q_1$ is of the order $\mathcal{O}(a^2)$ this will be true as long as $\Lambda$ is not really large and positive. This ensures that $1-\frac{\Lambda}{3}q_1 P_2(\cos\theta)>0$, so the square root is well defined. The function $k_2(r)$ approaches the constant $-q_1 \Lambda/6$ for large $r$. Loosely speaking requiring $\Lambda$ not to be too large ensure that the cosmological horizon is sufficiently far away from boundary of the slowly rotating fluid. In this case we can also be certain that our approximation remains valid within the fluid.

Next the conformally related metric is $\tilde{g}_{ab}=\tilde{\Omega}^2g_{ab}$, and we define $\mathcal{I}$ to be the boundary of our conformal metric at $r=\infty$, which coincides with $\tilde{\Omega}=0$. Now, $f(\theta)$ is non-vanishing and smooth, and hence it is easily shown that, similar to the case of Schwarzschild-(anti)-de Sitter space, $\tilde{\Omega}^2g_{ab}$ has the topology of $\mathbb{R} \times S^2 $ at $\tilde{\Omega}=0$. Moreover $\nabla_a \tilde{\Omega}$ is non-vanishing on $\mathcal{I}$. We already showed that the Weyl tensor identically vanishes on $\mathcal{I}$ and hence one verifies that the Hartle-Thorne-(anti)-de Sitter metric is always asymptotically (anti)-de Sitter. This is of course the expected result, but is in stark contrast to Hartle-Thorne metric without cosmological term as already mentioned before.

\section{The Wahlquist metric}
\label{sec:wahl}

In this section we review the Wahlquist metric in the slow rotating limit. The Wahlquist metric~\cite{Wahlquist:1968zz} is a rotating perfect fluid solution to Einstein's equation and is one of few known rotating interior solutions, with a problematic equation of state though. The metric can be written as
\begin{multline}
  ds^2=f(dt-\tilde{A}d\varphi)^2\\-r_0^2(\zeta^2+\xi^2)[\frac{d\zeta^2}{(1-\tilde{k}^2\zeta^2)\tilde{h}_1}+\frac{d\xi^2}{(1+\tilde{k}^2\xi^2)\tilde{h}_2}+\frac{\tilde{h}_1\tilde{h}_2}{\tilde{h}_1-\tilde{h}_2}d\varphi^2]
\end{multline}
where the functions $f(\zeta,\xi)$, $\tilde{h}_1(\zeta)$, and $\tilde{h}_2(\xi)$ are given by
\begin{align}
  f &= \frac{\tilde{h}_1-\tilde{h}_2}{\zeta^2+\xi^2}
  \tilde{A} = r_0\Bigl(
  \frac{\xi^2\tilde{h}_1+\zeta^2\tilde{h}_2}{\tilde{h}_1-\tilde{h}_2}
  -\xi_A^2\Bigr) \\
  \tilde{h}_1(\zeta) &= 1+\zeta^2+\frac{\zeta}{\kappa^2}
  \Bigl(
  \zeta_+\frac{1}{\tilde{k}}\sqrt{1-\tilde{k}^2\zeta^2}\arcsin(\tilde{k}\zeta)
  \Bigr) \\
  \tilde{h}_2(\xi) &= 1-\xi^2-\frac{\xi}{\kappa^2}
  \Big(\xi_-\frac{1}{\tilde{k}}\sqrt{1+\tilde{k}^2\xi^2}\sinh^{-1}(\tilde{k}\xi)
  \Bigr).
\end{align}
The constant $\xi_A$ is defined by $\tilde{h}_2(\xi_A)=0$. The pressure and density are given by
\begin{align}
  p&=\frac{1}{2}\mu_0(1-\kappa^2 f) + \Lambda\\
  \mu&=\frac{1}{2}\mu_0(3\kappa^2 f-1) - \Lambda.
\end{align}
and satisfy the unphysical equation of state
\begin{align}
  \mu + 3p = \frac{1}{2}\mu_0(3\kappa^2 f-1) - \Lambda + 
  \frac{3}{2}\mu_0(1-\kappa^2 f) + 3\Lambda 
  = \mu_0 + 2\Lambda
\end{align}
where $\mu_0$ is a constant.

The surface of zero pressure is given by the modified equation
\begin{align}
  \kappa^2 f=1+\frac{2\Lambda}{ \mu_{0} }
\end{align}
The non-rotating limit of the Wahlquist metric is given by the Whittaker metric
\begin{align}
  ds^2=-f_0dt^2+\frac{2}{\kappa^2\mu_{0} }
  \Bigl(
  \frac{dX^2}{f_0}+\sin^2X(d\Theta^2+\sin^2\Theta d\phi^2)
  \Bigr)
\end{align} 
where
\begin{align}
  f_0(X) = 1+\frac{1}{\kappa^2}(1-X \cot X).
\end{align}
The vanishing pressure surface is given by
\begin{align}
  X\cot X=\kappa^2-\frac{2 \Lambda}{ \mu_{0} }
\end{align}
and positivity of pressure in the interior implies
\begin{align}
  X\cot X>\kappa^2-\frac{2\Lambda}{ \mu_{0} }.
\end{align}
These imply we need $\kappa^2-\frac{2\Lambda}{ \mu_{0} }<1$. The coordinate transformations required to convert the Wahlquist metric into Hartle's coordinates were derived in~\cite{Bradley:1999gs}. Here we present the $l=2$ functions in Hartle's coordinates, and we refer the interested reader to~\cite{Bradley:1999gs} for the remaining components of the metric
\begin{align}
  h_2= &(x \cos x \frac{\sin^2 x+3\kappa^2-3}{6\sin^3 x}+\frac{\kappa^2}{6}\\&-\frac{1}{4}+\frac{3-6\kappa^2-2x^2}{12 \sin^2 x}+\frac{x^2}{4\sin^4 x})\frac{\mu_{0}r_0^2}{2\kappa^2}\\
  k_2= &(\frac{5x^2-3+3\kappa^2}{\sin^2 x}-2x^2+3-\kappa^2\\&-x\cos x\frac{3\kappa^2-6+5\sin^2x}{\sin^3 x}-\frac{3x^2}{\sin^4 x})\frac{\mu_{0} r_0^2}{12\kappa^2}.
\end{align}

\section{Matching conditions}
\label{sec:match}

It was shown in~\cite{Bradley:1999gs} that the Walhquist metric cannot be matched to an asymptotically flat vacuum exterior. In this section we attempt to match the slow rotating Wahlquist metric to the Hartle-Thorne-(anti)-de Sitter Lambda-vacuum exterior. In the interior fluid region, we match at the zero pressure surface. The matching conditions are
\begin{align}
  ds^2_{(V)}|_S &= ds^2_{(W)}|_S\\
  K_{(V)}|_S &= K_{(W)}|_S
\end{align}
where $K$ is the extrinsic curvature, and $S$ is the zero pressure surface. The subscripts $(W)$ and $(V)$ stand for Wahlquist and vacuum, respectively. Suitable hypersurfaces for matching in the vacuum exterior were given by Roos in~\cite{Roos:1976}, and are given by the condition
\begin{align}
  g_{\phi\phi}+\tilde{\Omega}g_{t\phi}+\tilde{\Omega}^2g_{tt} =
  1-\tilde{C} \label{Roos}
\end{align}
where $\tilde{\Omega}$ and $\tilde{C}$ are two constants. We note Roos' derivation of this still works in the presence of a non-zero cosmological constant. The matching conditions for slowly rotating fluids were derived in~\cite{Bradley:2006eg}. In the following, quantities evaluated at the matching surface will be denoted by a subscript~$1$. 

\subsection{Zero-th order matching}

To zero-order in the rotation parameter the exterior matching surface takes place on the history of the sphere $ \mathbb{R}\times S^{1}$ described by $r=r_1$. The Wahlquist metric to zero-th order in the rotation parameter is described by the Whittaker metric
\begin{align}
  ds^2=-f_0dt^2+\frac{2}{\kappa^2\mu_{0} }
  [\frac{dX^2}{f_0}+\sin^2X(d\Theta^2+\sin^2\Theta d\phi^2)]
\end{align} 
and the normal one-form has components
\begin{align}
  n_a^{(W)}=(0,1,0,0)\sqrt{g_{11}^{(W)}}
\end{align}
We rescale the interior time coordinate by $t\rightarrow c_4t$. The interior matching surface is given by the zero pressure surface, $x=x_1$ (not to be confused with the previous $x$ in Section~\ref{sec:num}) where $x_1$ is a solution to
\begin{align}
  x\cot x=\kappa^2-\frac{2\Lambda}{ \mu_0 }.
\end{align}  
Then solving the matching conditions on this zero pressure surface give us the following 
\begin{align}
  r_1 &= \frac{2^{1/2}}{\kappa\mu_{0}^{1/2}}\sin x_1\\
  c_4 &= \cos x_1\\
  M & =\frac{r_1}{2\kappa^2}(\kappa^2-\cos^2 x_1(1+\frac{2\Lambda}{ \mu_{0} })) -
  \frac{\Lambda}{6}r_1^3.
\end{align}

\subsection{First order matching}

To first order in the slow rotation parameter, the only change in the metrics will be the addition of  $(d\phi-\omega dt)^2 $ in place of $d\phi^2$. The matching surface is the same as it was to zero-th order. In the Wahlquist interior, we have
\begin{align}
  \omega=\frac{\mu_{0} r_0}{2\sin^2 x}(1-x \cot x),
\end{align}
and we apply a rigid rotation to the Wahlquist interior $\phi\rightarrow \phi+\Omega t$ where $\Omega$ is a constant. The matching equations give us the following conditions on the zero pressure surface
\begin{align}
  \omega^{(V)} = c_4(\omega^{(W)}-\Omega),\quad 
  \frac{d\omega^{(V)}}{dr} = c_4\frac{d\omega^{(W)}}{dx}\frac{dx}{dr},
\end{align}
which we can solve for the constants $\Omega$ and $a$
\begin{align}
  a &= \frac{r_0}{3\cos x_1}
  \frac{2x_1\cos^2x_1-3\sin x_1 \cos x_1+x_1}{x_1-\sin x_1 \cos x_1+\frac{4\Lambda \sin^2 x_1 \tan x_1}{3\mu_{0}} },\\
  \Omega &= \frac{x_1 r_0 \mu_{0}}{6\sin x_1\cos x_1}.
\end{align}

\subsection{Second order matching}

The pressure of the Wahlquist fluid up to second order in the rotation parameter was derived in~\cite{Bradley:1999gs}, and is of the form
\begin{align}
  p=p_0+p_{20}+p_{22}P_2(\cos\theta),
\end{align} 
where $p_0$ is the pressure to zero-th order. The equation of the zero pressure surface in the interior fluid region is therefore given by
\begin{align}
  x=x_1+r_0^2 \xi,
\end{align}
where
\begin{align}
  \xi=-[\frac{p_{20}+p_{22}P_2(\cos\theta)}{p_0'}]_{\mid x=x_1}.
\end{align}
In the vacuum exterior, the slow rotating matching surface will be given by
\begin{align}
  r=r_1+a^2 \chi.
\end{align}
Substituting this into~(\ref{Roos}), we obtain
\begin{align}
  \chi=\chi_0+\chi_2 P_2(\cos\theta).
\end{align}
We can apply a rescaling of the $t$ coordinate in the fluid region by $t\rightarrow c_4(1+r_0^2c_3)t$.

\begin{figure}
 \centering
 \includegraphics[width=0.48\textwidth,height=0.48\textwidth]{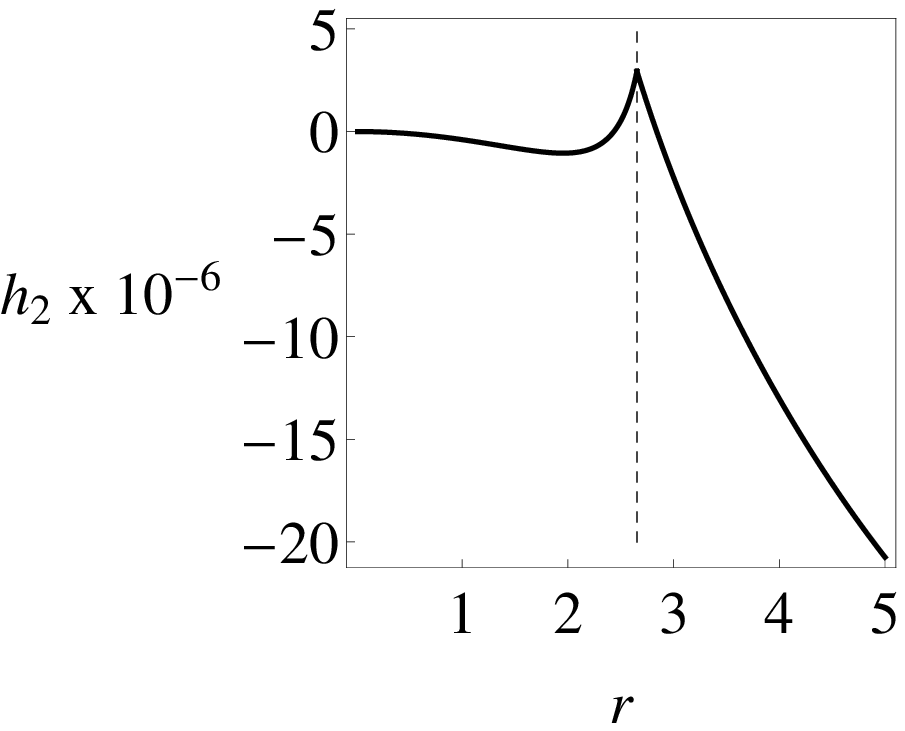}
 \hfill
 \includegraphics[width=0.45\textwidth,height=0.47\textwidth]{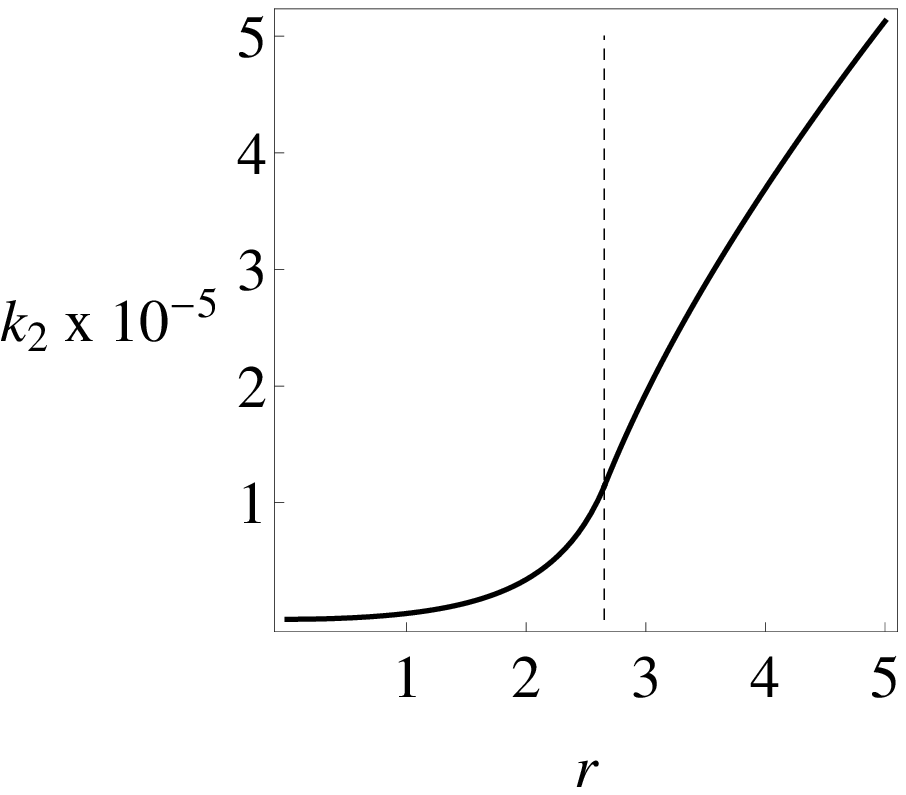}
 \caption{The result of numerically matching the functions $h_2$ and $k_2$ with parameter choices $\mu_0=1$, $\kappa=0.5$, $r_0=0.1$ and $\Lambda=-0.1$. The dashed line indicates the radius of the zero pressure surface at $r=2.65$. As can be seen from the left figure, the matching conditions only require these functions to be $\mathcal{C}^0$ on the matching surface. }
  \label{fig:match}
\end{figure}

The full second order matching conditions are given in~\cite{Bradley:2006eg}. All but the $l=2$ matching equations are easily solved for, and they follow the $\Lambda=0$ case almost identically. Two of the $l=2$ matching equations are given by
\begin{align}
  k_2^{(W)}(x_1)=k_2^{(V)}(r_1),
  \label{k equation} \\
  h_2^{(W)}(x_1)=h_2^{(V)}(r_1). 
  \label{h equation}
\end{align}
Once these are satisfied the remaining matching conditions are fulfilled automatically as a consequence of the field equations and the lower order matching conditions. However we don't know the functions $h_2$ and $k_2$ analytically in the vacuum region so  instead we find them numerically, using the procedure outlined in section~\ref{sec:num}. In order to find this numerical solution we need suitable boundary conditions, we choose $h_2^{(V)}(r_1)=h_2^{(W)}(x_1)$ and $k_2^{(V)}(r_1)=k_2^{(W)}(x_1)$, this ensures (\ref{h equation}) and (\ref{k equation}) are satisfied. We show this matching for a particular choice of parameter values in Fig.~\ref{fig:match} 
This shows that the Wahlquist fluid can in fact be matched to an asymptotically empty (with cosmological term) vacuum exterior up to second order in the rotation parameter. It was shown in~\cite{Bradley:2000sj} that the zero pressure surface of the Wahlquist metric is always prolate. On the other hand, one would expect an initially spherical body to becomes oblate when it starts rotating. Our result does not change these facts, however, it shows us that in some sense the cosmological constant is able to compensate for this and keep the fluid body in equilibrium. 

\section{Conclusions}
\label{sec:concl}

The motivation of this work is based two well-known results of slowly rotating fluids. Firstly, the exterior metric of a slowly rotating fluid cannot be match to an asymptotically flat Kerr spacetime. Secondly, it can be match if one drops the requirement of asymptotic flatness. This led us to the speculation that the inclusion of the cosmological constant, with its (anti)-de Sitter asymptotic behaviour, should make the matching somewhat more natural. In particular, inspection of~(\ref{Br2}) shows that the term which spoils asymptotic flatness is $q_1 r^2$, which is exactly of the form of the $\Lambda$ term in the Schwarzschild-de Sitter metric. 

It is probably not too surprising then that we were indeed able to match the Wahlquist fluid to an asymptotically empty (with cosmological term) vacuum exterior up to second order in the rotation parameter. This allows us to interpret the Wahlquist metric as an isolated rotating body, an interpretation that is not possible in the absence of the cosmological constant because the non-asymptotic flatness is unnatural.

Our results can be generalised along the lines of~\cite{Bradley:2006eg} and show that any slowly rotating perfect fluid with cosmological term can be matched to the Hartle-Thorne-(anti)-de Sitter exterior up to second order in the fluids angular velocity. It would also be interesting to include charge in this discussion and revisit the results of~\cite{Fodor:2002nt} with cosmological term. In our numerical studies we have restricted ourselves to negative $\Lambda$ in order to avoid integrating through the cosmological horizon. It would be very interesting to extend our work along these lines to better understand the exact effects a positive cosmological constant would have on slowly rotating fluid, in particular as all observational evidence points towards a small positive value.

\subsection*{Acknowledgement}

We are very grateful to Gyula Fodor for valuable comments on the manuscript.

\end{document}